\documentstyle[12pt]{article}

\newcommand{\beq}{\begin{equation}}
\newcommand{\eeq}{\end{equation}}

\renewcommand{\theequation}{\thesection.\arabic{equation}}
\newcounter{ehsctr}

\newcommand{\NPB}[1]{{\it Nucl. Phys.}\ {\bf B{#1}}}
\newcommand{\PLB}[1]{{\it Phys. Lett.}\ {\bf B{#1}}}
\newcommand{\PRD}[1]{{\it Phys. Rev.}\ {\bf D{#1}}}
\newcommand{\PRL}[1]{{\it Phys. Rev. Lett.}\ {\bf #1}}
\newcommand{\NCA}[1]{{\it Nuovo Cim.}\ {\bf {#1}A}}

\newcommand{\hc}{ {\rm h.c.} }
\newcommand{\ME}{ M_{ETC} }
\newcommand{\gE}{ g_{ETC} }

\newcommand{\ETC}{ extended technicolor }

\newcommand{\Zbb}{ {Zb{\bar b}} }
\newcommand{\sw}{ s_\theta }
\newcommand{\cw}{ c_\theta }
\newcommand{\esc}{ {e\over\sw\cw} }
\newcommand{\half}{{1 \over 2}}

\begin{document}

\title{%
Walking Technicolor and the $Zb\bar b$ Vertex}

\author{%
R.S. Chivukula$^{a}$, E. Gates$^{b}$, E.H. Simmons$^{c}$
and J. Terning$^{d}$}

\date{}

\begin{titlepage}
\def\thepage {}        

\maketitle

\begin{center}
$^a$ Department of Physics, Boston University, Boston, MA  02215\\
$^b$ Enrico Fermi Institute, University of Chicago, Chicago, IL 60637\\
$^c$ Lyman Laboratory of Physics, Harvard University, Cambridge MA 02138\\
$^d$ Department of Physics, Yale University, New Haven, CT 06511\\
\end{center}

\bigskip
\begin{abstract}

A slowly-running technicolor coupling will affect
the size of non- oblique
corrections to the $\Zbb $ vertex from
extended technicolor dynamics.  We
show that while ``walking technicolor"
reduces the magnitude of the
corrections, they generally remain large enough
to be seen at LEP.

\end{abstract}
\bigskip
\bigskip
\begin{flushleft}
{\rm BUHEP-93-11 \\ EFI-93-27 \\ HUTP-93/A012 \\ YCTP-P10-93
\\ hep-ph/9305232 }
\end{flushleft}
\bigskip
\bigskip

\footnotetext[0]{Electronic mail addresses: sekhar@weyl.bu.edu,
gates@tyrone.uchicago.edu, simmons@physics.harvard.edu,
terning@genesis1.physics.yale.edu}

\end{titlepage}

\section{Introduction}
\label{sec:intro}

The origin of the diverse masses and mixings of the quarks and leptons
remains a mystery; most puzzling is the origin of the top quark's large
mass.  In technicolor models \cite{technicolor}, the large top mass is
thought to arise from extended technicolor \cite{etc}  (ETC) dynamics at
relatively low energy scales\footnote{So long as no additional light
scalars couple to ordinary and techni-fermions \cite{scal,hscal}.}.
Recent work \cite{zbbar} has shown  that the  dynamics responsible for
generating the large top quark mass in extended technicolor models will
produce potentially large ``non-oblique" \cite{oblique} effects at the
$Zb\bar b$ vertex \footnote{In contrast, the $\Zbb$ effects in models with
additional light scalars (e.g. strongly-coupled ETC models) are
indistinguishable from those in the standard model \cite{zbbar}.}.  In this
note, we discuss what happens to these effects if the technicolor beta
function is  assumed to walk \cite{walk}.  We show that the size of the
signal is reduced but that it remains quite visible at LEP for many models.

\section{ETC's Effect on the $\Zbb$ vertex}
\label{sec:vertex}
\setcounter{equation}{0}

We begin by reviewing the results of ref. \cite{zbbar}.
Consider a model in which $m_t$ is generated by the exchange of  a
weak-singlet \ETC gauge boson of mass $M_{ETC}$ coupling  with
strength $\gE$ to the current
\begin{eqnarray} {\xi} {\bar\psi^i}_L \gamma^\mu T_L^{ik}
&+& {\xi'} {\bar t_R} \gamma^\mu U_R^k \\
\psi_L\ =\ \pmatrix{t \cr b \cr}_L\ \ &\ &\ \ \
T_L\ =\ \pmatrix{U \cr D \cr}_L \nonumber
\label{tmasscur}
\end{eqnarray}
where $U$ and $D$ are technifermions, $i$ and $k$ are weak and  technicolor
indices, and the coefficients $\xi$ and $\xi'$ are \ETC Clebschs expected
to be of order one.  At energies below $\ME$, ETC  gauge boson exchange
may be approximated by local four-fermion operators.   For example, $m_t$
arises from an operator coupling the  left- and right-handed pieces of the
current in Eq. (\ref{tmasscur})
\beq
   -\xi\xi' {\gE^2 \over  \ME^2}  \left({\bar\psi}_L^i \gamma^\mu
T_L^{iw}\right) \left( {\bar U^w}_R \gamma_\mu t_R \right) + \hc\ .
\label{topff}
\eeq
When this is Fierzed into a product of technicolor singlet densities, it is
seen to generate a mass for the top quark after the technifermions' chiral
symmetry breaking. We can use the rules of naive  dimensional analysis
\cite{ndaref} to estimate the size of $m_t$ generated by Eq. (\ref{topff}).
Assuming, for simplicity, that there is only doublet of technifermions and
that  technicolor respects an $SU(2)_L \times SU(2)_R$ chiral symmetry (so
that the technipion decay constant, $F$, is $v\approx 250$ GeV) we have
\beq
   m_t\ = \xi\xi' {\gE^2 \over \ME^2}
   \langle{\bar U}U\rangle\ \approx\ \xi\xi'{\gE^2 \over \ME^2} (4\pi v^3)\ .
\label{topmass}
\eeq

In the same language, we can also show that the \ETC boson responsible for
producing $m_t$ affects the $\Zbb$ vertex.  Consider the four-fermion
operator arising purely from the left-handed part of the current
(\ref{tmasscur}) -- the only part containing $b$ quarks.
\beq
  -\xi^2 {\gE^2\over\ME^2} \left({\bar \psi^i}_L \gamma^\mu
  T^{iw}_L \right) \left({\bar T^{jw}}_L \gamma_\mu \psi^j_L \right)\ .
\label{unfierz}
\eeq
When Fierzed into a product of technicolor singlet currents, this
includes\footnote{The Fierzed form of (\ref{unfierz})
also includes operators that are products of weak-singlet
left-handed currents; these will not affect the $\Zbb$ coupling.}
\beq
  -{\xi^2\over 2} {\gE^2\over
  \ME^2}\left({\bar\psi}_L\gamma^\mu\tau^a\psi_L \right) \left({\bar T}_L
  \gamma_\mu \tau^a T_L \right)\ ,
\label{fourferm}
\eeq
where the $\tau^a$ are weak isospin Pauli matrices.  As shown
in \cite{zbbar} this alters the $Z$-boson's tree-level coupling
to left-handed bottom quarks $g_L = \esc(-\half + {1\over 3}\sw^2)$ by
\renewcommand{\theequation}{\thesection.\arabic{equation}\alph{ehsctr}}
\addtocounter{ehsctr}{1}
\begin{eqnarray}
\delta g_L &=& -{\xi^2 \over 2} {\gE^2 v^2\over\ME^2} \esc(I_3)
\label{ta}\\
\addtocounter{ehsctr}{1}
\addtocounter{equation}{-1}
	&=& {1\over 4} {\xi\over {\xi'}} {m_t\over{4\pi v}}
\cdot \esc \label{tb}
\end{eqnarray}
\renewcommand{\theequation}{\thesection.\arabic{equation}}
Here eq. (\ref{tb}) follows from applying eq. (\ref{topmass}) to eq.
(\ref{ta}).

\section{Measuring the Effect at LEP}
\label{sec:measure}
\setcounter{equation}{0}

We now consider how best to experimentally measure the shift in  $g_L$
caused by extended technicolor.  Altering the $Zb\bar b$ coupling will
affect the  decay width of the $Z$ boson into $b$ quarks.  In addition,
there are flavor universal (oblique) corrections to the width, coming from
both technicolor and extended technicolor interactions.  At one loop, the
decay width of the  $Z$ is of the form
\beq
\Gamma_b^{corr.} \equiv \Gamma(Z \to b\bar b) =
(1 + \Delta\Gamma)(\Gamma_b + \delta\Gamma_b)\label{gammaa}
\eeq
where $\Gamma_b$ is the tree-level decay width, $\Delta\Gamma$ represents
the oblique corrections and $\delta\Gamma_b$ represents the non-oblique
(flavor-dependent) corrections.  We will refer to the non-oblique effect
of (\ref{ta}) on the decay width as $\delta\Gamma_b^{ETC}$.  Ratios of $Z$
decay widths into different final states are particularly sensitive to such
effects; we suggest studying the
ratio\footnote{This is simply related to the ratio
$\Gamma_b / \Gamma_h$ discussed in \cite{zbbar} :
\beq
\frac{\Gamma_b}{\Gamma_{h\neq b}} = \frac{\Gamma_b/\Gamma_h}{1 -
\Gamma_b/\Gamma_h \nonumber}
\eeq
but is more convenient to work with.  We thank A. Pich for pointing this
out.}
of the $Z$ decay width into $b\bar b$ and the $Z$ decay width  into all
non-$b\bar b$ hadronic final states:  $\Gamma_b / \Gamma_{h\ne b}$.  This
is accessible to the current LEP experiments.

This particular ratio has several features  to recommend it. First, since
it is a ratio of hadronic widths, the leading QCD corrections cancel in the
limit of small quark masses.  Second,  eq. (\ref{gammaa}) implies that the
fractional change in this particular ratio is approximately the fractional
shift in $\Gamma_b$:
\beq
\Delta_R \equiv \frac{\delta {\left(\Gamma_b / \Gamma_{h \neq b}\right)}}
{\left(\Gamma_b / \Gamma_{h \neq b} \right)}
\approx \frac{\delta\Gamma_b}{\Gamma_b}.
\eeq
This is easily related to the change in $g_L$ that \ETC effects cause:
\beq
\Delta_R^{ETC} \approx \frac{\delta\Gamma_b^{ETC}}{\Gamma_b}
\approx \frac{2 g_L \delta g_L}{g_L^2 + g_R^2}.
\label{dell}
\eeq
For our benchmark ETC model with two technifermion flavors,
\beq
\Delta_R^{ETC} \approx - 3.7\%\cdot \xi^2\cdot
\left(\frac{m_t}{100{\rm GeV}}\right).
\eeq

There is also a fractional shift in $\Gamma_b$ arising from 1-loop diagrams
involving longitudinal $W$-boson exchange and internal top quarks.  This
has already been calculated \cite{loopwt}; it is of order  -0.7\% (-2.5\%)
for $m_t$ = 100 (200) GeV.   This source of corrections to $\Gamma_b$
(which we shall call $\Delta^W_R$)  occurs both in the standard model
(where it is the dominant non-oblique correction to $\Gamma_b$) and in \ETC
models. Note that both $\Delta_R^W$ and $\Delta_R^{ETC}$ act to {\it
decrease} $\delta\Gamma_b / \delta\Gamma_{h\neq b}$. Then in comparing the
size of $\Delta_R$ in the standard model with that in ETC models, we are
comparing $\Delta_R^W$ to $\Delta_R^W + \Delta_R^{ETC}$. The expected LEP
precision of 2.5\% for measurement of $\Delta_R$ \cite{zbblep}  should
suffice to distinguish them.

\section{Walking Technicolor}
\label{sec:betafn}
\setcounter{equation}{0}

The dimensional estimates employed in section  \ref{sec:vertex} are
self-consistent so long as the \ETC interactions may be treated as a small
perturbation on the technicolor dynamics, i.e. so long as $\gE^2 v^2/\ME^2
< 1$ and there is no fine-tuning \cite{scal}.  Note that the rules of naive
dimensional analysis do not require that $\ME$ be large, only that $\gE^2
v^2/\ME^2$ (or equivalently $m_t/4\pi v$) be small. However, these
estimates (in particular, the relationship \ref{topmass}  between
$\left(\frac{g^2 v^2}{\ME^2}\right)$ and $\left(\frac{m_t}{4\pi v}\right)$)
are typically modified in ``walking technicolor'' models \cite{walk}  where
there is an enhancement of operators of the form  (\ref{topff}) due to a
large anomalous dimension of the technifermion mass operator.

Let us define what is meant by a ``walking'' technicolor coupling. The
beta function for an $SU(N)$ technicolor force has the same form as
that for QCD.  At leading order it is simply
\beq
\beta(\alpha_{TC}) = - b\ \alpha_{TC}^2\ + \ {\cal O}(\alpha_{TC}^3)
\eeq
where (for technifermions in the fundamental representation)
$b$ is related to the technicolor group and the number of technifermion
flavors ($n_f$) by
\beq
b = \frac{1}{2 \pi} \left( \frac{11}{3} N - \frac{2}{3} n_f \right)\ .
\label{bbeqn}
\eeq
For our benchmark model with two technifermion flavors, setting $N$ =
2 yields $b = \frac{3}{\pi}$.   Adding more flavors of technifermions to
the model decreases $b$ so the TC coupling falls off
relatively slowly with increasing momentum scale (it ``walks'').

The expected effect of a walking technicolor coupling on $\Delta_R^{ETC}$
can be outlined fairly briefly.  When the technicolor coupling becomes
strong and the  technifermion condensate $\langle \bar T T \rangle$ forms,
a dynamical mass $\Sigma(p)$ is also generated for the technifermions.  As
discussed in ref. \cite{walk},  having $\alpha(p)$ fall off slowly with
increasing $p$ causes $\Sigma(p)$ to decrease more slowly with rising $p$
than it would in a `running' TC theory.  Since the  technifermion
condensate is
\beq
\langle \bar T T \rangle \sim \int\limits^{M_{ETC}^2}_0 dk^2 \Sigma(k),
\eeq
enhancing $\Sigma$ increases $\langle \bar T T \rangle$.  According to eq.
(\ref{topmass}) this means  that a walking TC coupling increases $m_t$ for
a given ETC scale $\ME$. The factor $\left(\frac{g^2 v^2}{\ME^2}\right)$
appearing in our expression (\ref{ta}) for $\delta g_L$ is therefore
smaller than $\left(\frac{m_t}{4\pi v}\right)$ in an ETC model with
walking TC. Thus, the expected size of $\Delta_R^{ETC}$ is reduced.

\section{Numerical Results}
\label{sec:dse}
\setcounter{equation}{0}

To illustrate the effect of walking technicolor  on the size of
$\Delta_R^{ETC}$, we have studied coupled ladder-approximation
Dyson-Schwinger equations  \cite{walk} for the dynamical technifermion and
top quark masses, $\Sigma(p)$ and $m_t(p)$.   The gap equations always
possess a chiral symmetry preserving solution  with $m_t$ and $\Sigma$ both
equal to zero.  Our interest is in finding  chiral symmetry {\it violating}
solutions with both $m_t$ and  $\Sigma$ non-zero. We have focused
on $SU(N+1)_{ETC} \to SU(N)_{TC}$ models with a full family of
technifermions.

In Landau gauge and after the angular integrations have been performed,
we approximate the gap equations by \cite{aeqns}
\begin{eqnarray}
\Sigma(p) & = &\ C_{2}^{TC} \int\limits^{\infty}_0
\frac{3 \alpha_{TC}(M[p,k])}{\pi M[p^2,k^2]}
\frac{\Sigma(k)}{k^2 + \Sigma^2(k)} k^2 dk^2 \nonumber \\
 & + &\ c_1 \int\limits^{\infty}_0
\frac{3 \alpha_{TC}(M[p,k,M_{ETC}])}{\pi M[p^2,k^2,M_{ETC}^2]}
\frac{\Sigma(k)}{k^2 + \Sigma^2(k)} k^2 dk^2 \nonumber \\
 & + &\ c_2 \int\limits^{\infty}_0
\frac{3 \alpha_{TC}(M[p,k,M_{ETC}])}{\pi M[p^2,k^2,M_{ETC}^2]}
\frac{m_t}{k^2 + m_t^2} k^2 dk^2 \nonumber \\
 & + &\ C_{2}^{QCD}  \int\limits^{\infty}_0
\frac{3 \alpha_{QCD}(M[p,k,M_{ETC}])}{\pi M[p^2,k^2,M_{ETC}^2]}
\frac{\Sigma(k)}{k^2 + \Sigma^2(k)} k^2 dk^2  \\
&\ & \nonumber \\
m_t & = &\ c_3 \int\limits^{\infty}_0
\frac{3 \alpha_{TC}(M[p,k,M_{ETC}])}{\pi M[p^2,k^2,M_{ETC}^2]}
\frac{m_t}{k^2 + m_t^2} k^2 dk^2 \nonumber \\
 & + &\ c_4 \int\limits^{\infty}_0
\frac{3 \alpha_{TC}(M[p,k,M_{ETC}])}{\pi M[p^2,k^2,M_{ETC}^2]}
\frac{\Sigma(k)}{k^2 + \Sigma(k)^2} k^2 dk^2 \nonumber \\
 & + &\ C_{2}^{QCD}  \int\limits^{\infty}_0
\frac{3 \alpha_{QCD}(M[p,k,M_{ETC}])}{\pi M[p^2,k^2,M_{ETC}^2]}
\frac{\Sigma(k)}{k^2 + \Sigma^2(k)} k^2 dk^2
\end{eqnarray}
where $M[x,y]$ signifies the greater of $x$ and $y$, $C_{2}^{TC} =
\frac{N^2-1}{2N}$, $C_{2}^{QCD} = \frac{4}{3}$ and the coefficients
$c_i$ are
\beq
c_1 = \frac{1}{2 N (N+1)} \ \ \
c_2 = \frac{1}{2}\ \ \
c_3 = \frac{N}{2 (N+1)}\ \ \
c_4 = \frac{N}{2}
\eeq
We have ignored the mass splittings between the extended technicolor gauge
bosons and used $M_{ETC}$ to stand for the masses of all the
heavy \ETC gauge bosons in the gap equations.

To integrate the gap equations, we use the following 1-loop form for
the running of the technicolor coupling:
\begin{eqnarray}
\alpha_{TC}(p)& = & 2 \alpha^c_{TC}\ \ \ \ \ \ \ \ \ \ \ \ \ \ \ \ \ \
p < \Lambda_c \nonumber\\
& = & {{2 \alpha^c_{TC}}\over{1+ b \alpha^c_{TC}
\ln\left({{p^2}\over{\Lambda_c^2}}\right)}} \ \ \ p \geq \Lambda_c
\end{eqnarray}
where $\alpha^c_{TC} \equiv \pi / 3 C_{2}^{TC} $  is the `critical' value
for chiral symmetry breaking, and $\Lambda_c$ is to be determined by
requiring that the model reproduce the correct electroweak gauge boson
masses.  At energies below the \ETC scale $M_{ETC}$, the beta-function
parameter $b$ is given by eq. (\ref{bbeqn}); above $M_{ETC}$, it is
\begin{eqnarray}
b_{ETC} &=& \frac{1}{2 \pi} \left( \frac{11}{3} (N + 1)
- \frac{2}{3} n_f \right)\nonumber\\
&=& b+{1\over{6 \pi}}\ .
\end{eqnarray}

We set the scale of the chiral symmetry breaking and the dynamical
masses by using the calculated $\Sigma(p)$ to compute the technipion decay
constant \cite{PS}
\beq
f^2 = \frac{N_{TC}}{16\pi^2} \int\limits^\infty_0
 \frac{4 k^2 \Sigma^2 + \Sigma^4}{(k^2 + \Sigma^2)^2} dk^2.
\eeq
In one-family technicolor models, $f \approx 125$ GeV.

For a given value of $b$, we vary the ETC scale, $M_{ETC}$, until we
obtain a chiral symmetry violating solution to the gap equations with a
particular value of $m_t$.  Knowing $M_{ETC}$ allows us to use equations
(\ref{ta}) and (\ref{dell}) to find the value of $\Delta_R^{ETC}$
associated with our initial values of $b$ and $m_t$. In applying (\ref{ta})
we recall that $g^2_{ETC} \equiv 4 \pi \alpha_{TC}(M_{ETC})$ and we set
$\xi = \xi' = \frac{1}{\sqrt 2}$ as is appropriate for our ETC models.

Our numerical results for an $SU(3)_{ETC} \to SU(2)_{TC}$ model are shown
in fig. \ref{fig:ntctwo}.  Here, $\Delta_R^{ETC}$ is plotted as a function
of $A \equiv (b\alpha^c_{TC})^{-1}$ for several values of the top quark
mass.  Similar results for an $SU(5)_{ETC} \to
SU(4)_{TC}$ model are plotted in fig. \ref{fig:ntcfour}.  In the small-A
(``running") regime of the plots, $\Delta_R^{ETC}$ is of order a few
percent, in good agreement with the estimates from naive dimensional
analysis.   As one moves towards the large-A (``walking") regime, the  size
of the effect decreases as we expected.  Note that the decrease is {\it
very} gradual. For the large top quark masses shown,  $\Delta_R^{ETC}$
generally remains big enough to be visible at LEP even if the TC coupling
runs very slowly.

Fig. \ref{fig:ntcmulti} compares the variation of $\Delta_R^{ETC}$ with $A$
found for several $SU(N+1)_{ETC} \to SU(N)_{TC}$ models with $m_t$ set to
140 GeV.  Note that the size of $\Delta_R^{ETC}$ grows with $N$ and
that $\Delta_R^{ETC}$ depends much less strongly on $A$ as $N$ increases.

\section{Conclusions}
\label{sec:concl}

In this note, we have discussed the degree to which extended technicolor
effects reduce the $Zb\bar b$ coupling in models with a walking technicolor
beta-function.  We chose the variable $\Delta_R$ (fractional shift in the
ratio of $Z$ hadronic widths $\Gamma_b / \Gamma_{h\ne b}$) as most suitable
for  measurement of the shift in the coupling.   We indicated why one
expects models with a slowly running technicolor beta function to have a
smaller $\Delta_R^{ETC}$ than models with a running TC beta function.  Then
we presented a numerical analysis of dynamical chiral symmetry breaking to
illustrate how strongly the technicolor beta function affects the size of
$\Delta_R^{ETC}$.  Our results show that while $\Delta_R^{ETC}$ is reduced
in walking technicolor models, it generally remains large enough to be
visible at LEP.


\centerline{\bf Acknowledgments}

We thank B. Holdom, A. Pich and T. Appelquist for helpful conversations.
We appreciate the hospitality of the Aspen Center for Physics, where
part of this work was completed.
R.S.C. and J.T. each acknowledge the support of a Superconducting
Super Collider National Fellowship from the Texas National
Research Laboratory Commission.  R.S.C. also acknowledges the support
of an Alfred P. Sloan Foundation Fellowship, an NSF Presidential
Young Investigator Award and a DOE Outstanding Junior Investigator Award.
The work of E.G. is supported in part by the Department of Energy (at the
University of Chicago and Fermilab and by National Aeronautics and Space
Administration through NAGW-2381 (at Fermilab).
{\it This work was  supported in part by the National Science Foundation
under grants PHY-9218167 and PHY-9057173, by the Department of Energy under
contract DE-FG02-91ER40676, by the Texas National Research Laboratory
Commission under grants RGFY93-278B and RGFY92B6, and by the Natural
Sciences and Engineering Research Council of Canada.}


\begin{figure}[p]
\caption{Plot of $\Delta_R^{ETC}$ as a function of walking parameter A in
an $SU(3)_{ETC} \to SU(2)_{TC}$ model. The dotted (solid, dashed) curve is
for a top quark mass of 100 (140, 180) GeV.}
\label{fig:ntctwo}
\end{figure}

\begin{figure}[p]
\caption{Plot of $\Delta_R^{ETC}$ as a function of walking parameter A in
an $SU(5)_{ETC} \to SU(4)_{TC}$ model. The dotted (solid, dashed) curve is
for a top quark mass of 100 (140, 180) GeV.}
\label{fig:ntcfour}
\end{figure}

\begin{figure}[p]
\caption{Plot of $\Delta_R^{ETC}$ as a function of walking parameter A
for $m_t = 140$ GeV in several $SU(N+1)_{ETC} \to SU(N)_{TC}$ models.}
\label{fig:ntcmulti}
\end{figure}

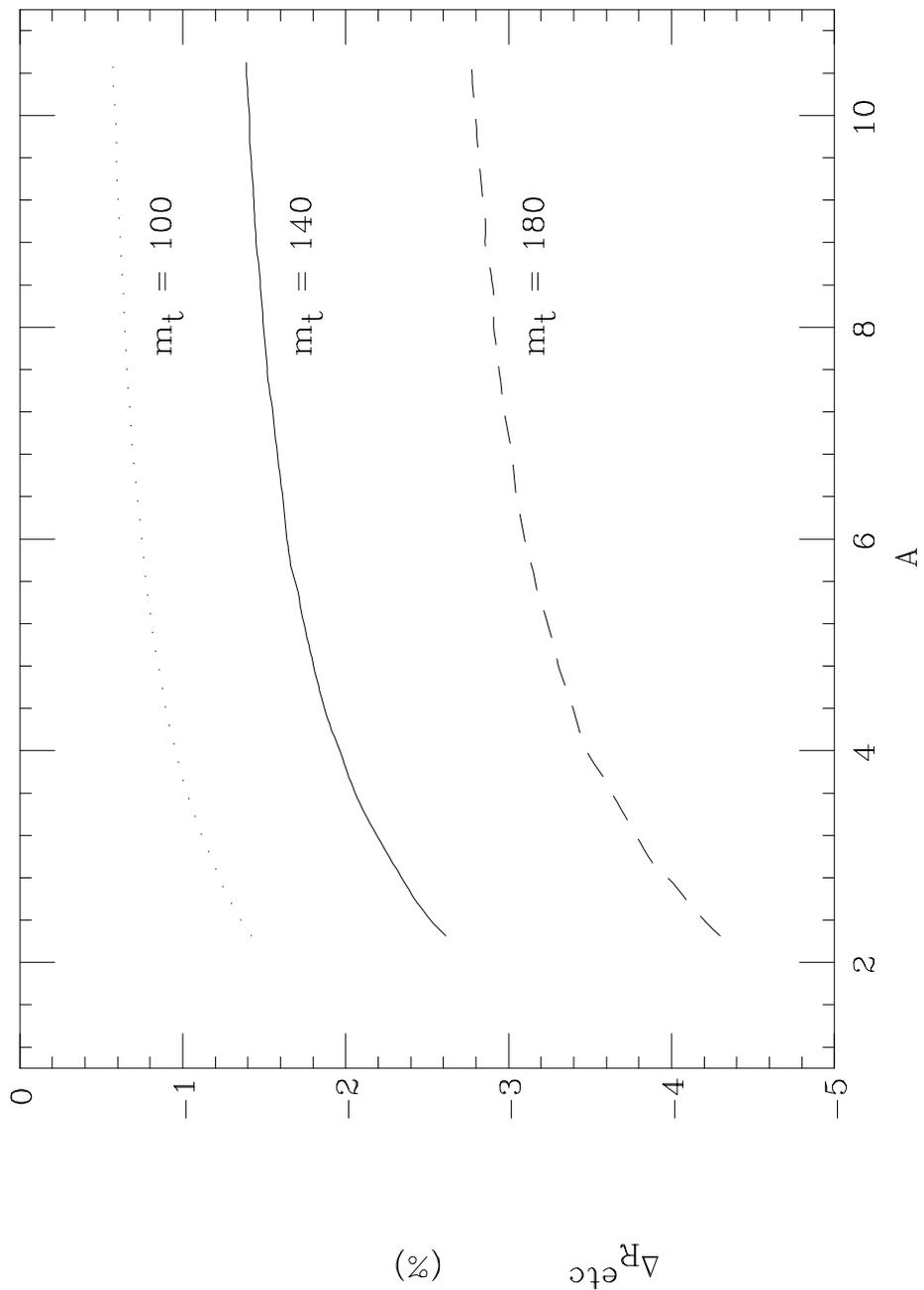
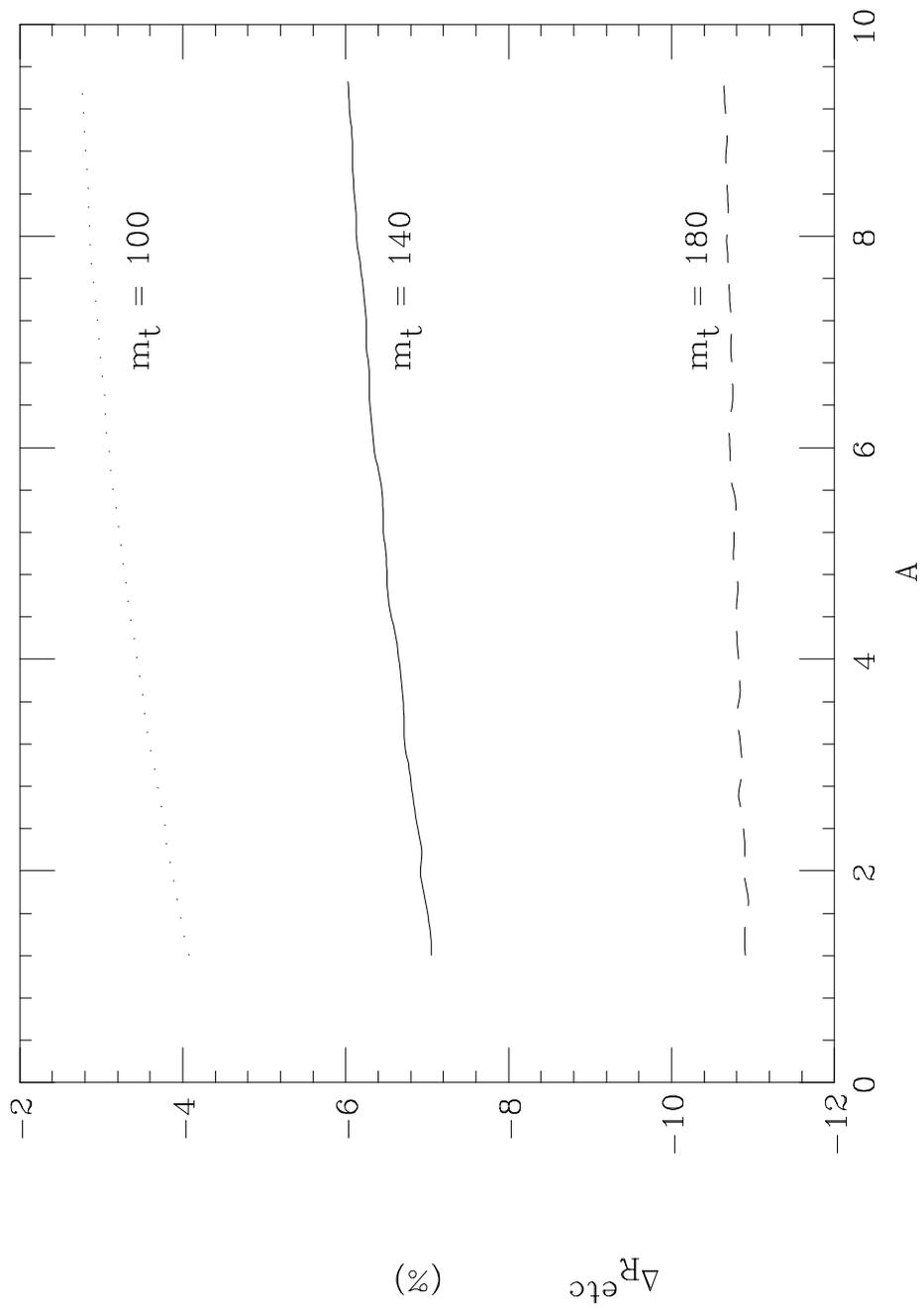
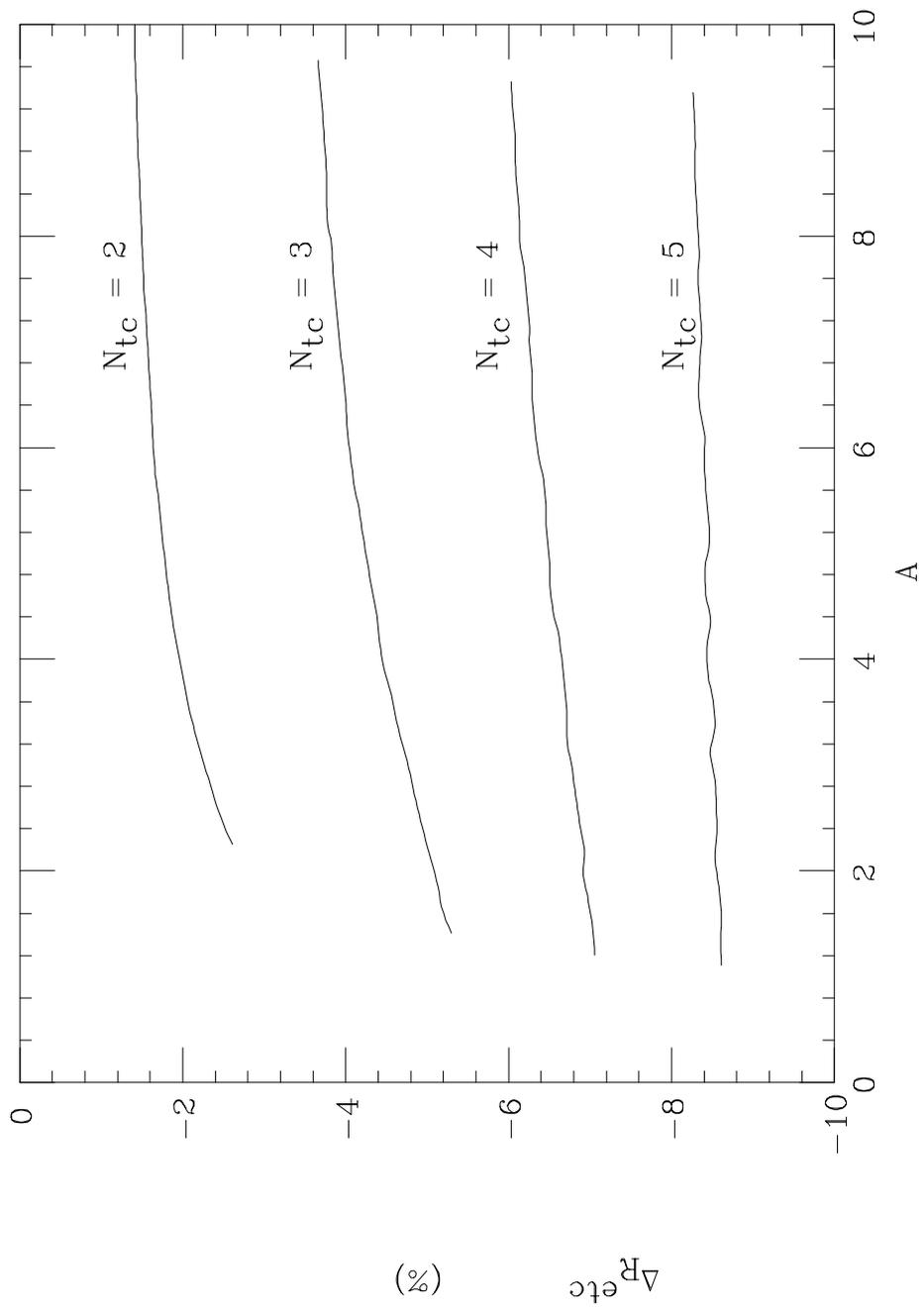
\end{document}